# Software Architecture and Subclassing Technique for Semiconductor Manufacturing Machines

[1]HyungTae Kim, [2]HaeJeong Yang
[*1,] Smart System Research Group, KITECH, CheonAn, South Korea
[2,] Mechanical Design Engineering, Korea Polytechnique University, SiHeung, South Korea
Email: [1]htkim@kitech.re.kr, [2]yhj@kpu.ac.kr

*Abstract.* **This paper proposed software architecture for operating an automatic semiconductor manufacturing machine. Recent machines for semiconductor process are required for high level of automation which are composed of motion control, machine vision, data acquisition and networking. These functions are executed through industrial equipments that are generally installed in a computer. The equipments occupy a great part of system resource and generate a lot of computation, so the software structure should be designed for efficiency. The proposed architecture is consisted of four layers and virtual equipments(VEs). The VEs are made by subclassing the physical equipments(PEs) and the layers are coded into thread which updates the status of VEs. Subroutines in a program refer to the pointer of VEs, and direct access to physical equipment are prohibited. The number of access (NOA) to PEs in typical industrial application was simulated for the unlimited access structure and the presented structure. The result showed that the proposed structure was more efficient than typical ones and irrespective of subroutines. This architecture was also applied to design a machine operating software and performed automatic wafer dicing.**

* Corresponding Author:
HyungTae Kim,
Smart System Research Group,
Korea Institute of Industrial Technology CheonAn, South Korea,
Email: mailto:htkim@kitech.re.kr    Tel:+82-41-589-8478

## 1. Introduction

Semiconductor manufacturing machines have been automated with intelligent peripherals and recent technologies. The conventional functions of the machines are integrated with precise positioning, vision inspection, signal analysis, and computer integrated manufacturing (CIM) protocol. These functions can be programmed with industrial boards such as motion controllers (MCs), framegrabbers (FGs), data acquisition (DAQ) boards, and network cards as shown in Fig. 1. These equipments are installed in PC and defined as physical equipments (PEs). Increase of machine intelligence heightens the complexity of program and the loads on computer system. But the access to PEs should be limited because PEs commonly perform numerical computation and have heavy processing load. Unlimited and random access from arbitrary subroutines to PEs burdens computer resource with overload and results in PC system failure or lag. So, when we design and develop an industrial application, program structure must be considered in terms of effective interface with PEs and minimize the access. Following researches are related with software structure for industrial machines.

Bogdanchinov developed DAQ software for SND detector [1]. He proposed web-base interface architecture for monitoring signals. The architecture was aimed at DAQ and data basing and three layers were used to handle raw events. Abstract layer deals with events, system layer handles file system and hardware layer supports hardware. Abdel-Samad developed an automatic control system for vacuum installations [2]. The valves, pumps and sensors were connected to programmable logic controller (PLC). The status of PLC was monitored on PC program. The detailed architecture was shown in his study, however the software seems to simply interface with an operator and communicate with the PLC. Oliveira researched a control system for a hazardous environment like Tokamak experiment, which is isolated from computers [3]. Several computers were connected on the network where data streams are transferred. The devices for monitoring the hazardous area were organized by fiber optics LAN. Wang presented an COM-base structure for sensor interfacing system and four levels of signal processing [4]. His architecture had four layers of sensor driver level, logical sensor level, fusion unit level and task unit level. The basic layer detected signals from external sensors, and higher logics were formed by the status on lower layers. The actions of monitoring devices were coded by COM in API level. Cruza proposed a software architecture for industrial robot control [5]. His structure was composed of hardware platform which must be processed for real time signal and higher level software which supervised the hardware and network. Anjos developed a software for robot manipulation in air craft industry using LABVIEW [6]. His architecture was based on discrete events but it was not general structure and can be used for specific targets. Song proposed a software architecture for operating multiple robots [7]. His architecture had four layers of motion control, behavior control, behavior planning and task coordination. The architecture was applied to steel-cutting machine which interfaced with a framegrabber, a PLC module and motion controllers. Recent researches on software architecture for robots approach to intelligence.[8] Bin developed reusable components for operating software of computer numerical control (CNC) machines [9]. He analyzed the actions of CNC during operation and proposed open CNC. The reusable components were used to construct a CNC application. An operating software was constructed for a lathe by open architecture programming [10]. The machine interfaced with a motion controller, an FPGA board for vibration and current analysis and a PLC. However, the software was built on the basis of events rather than architecture. A software was designed for detecting fault and failures during cutting metal [11]. The software gathers information about a CNC machine in real time and gave feedback to spindle. The software monitored the cutting process, but did not operate the machine. Zhang developed USB-based architecture for CNC [12]. PC is a host controller and communicates with spindles, and PLCs through USB. The architecture has several real time modules which are occupied by multi-thread.

As the level of automation increases in manufacturing machines, the number of signals which are obtained by sensors also increases. So, a machine interfaces with multiple signal boards, which makes software logics more complex. A main control PC in the machine has finite resources, but the operating software becomes slower and shutdown sometimes if the resource is concentrated in one of the signal boards. Therefore, it is important to distribute the resource to decrease the potentials of a software malfunction. This paper proposes a software structure for interfacing industrial peripherals and subclassing techniques for manufacturing machines. The peripherals are defined as PEs and coded into VEs. The VEs have status flags (SFs) and access functions which have command sets. VEs are allotted as pointers in a memory, which all subroutines refer to the pointers. Individual subroutines can access to PEs only through the pointers of VEs. The basic layer (BL) and logic layer (LL) update the SFs. Display layer (DL) shows current states to a user and process layer (PL) executes manufacturing process. The four layers are driven by window thread. To show efficiency of the presented structure, the number of access to PEs were checked for the typical and the proposed structure. This architecture was applied to operating software for an automated machine in the wafer dicing process.

## 2. Virtual equipment and subclassing
### 2.1 Motion control
The major function of a MC is to drive kinematic movement and maintain position in a machine. PC-MC suppliers provide API command sets to access the boards for programmers. Each MC has typical grammatical commands, but the conventional functions can be generalized. Common MCs have several functions like Fig. 2. These functions can be assorted into commands and status. The commands activate the functions of MCs such as S-curve/trapzoidal positioning, interpolation, stop, homing, jog, MPG and etc. The status shows current situation of movement like position, velocity, ± HW/SW limits, in-position, emergency, drive fault and etc. These features can be coded into virtual motion equipment(VME) using C++ class. Member functions of VME have general name space for motion control and symbolize the commands.

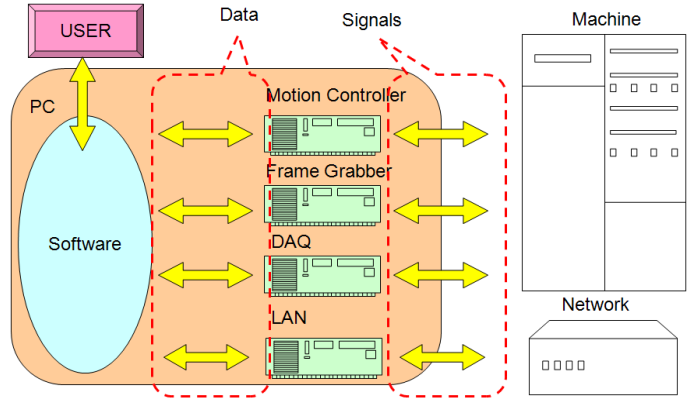

**Figure 1.** Interfacing concept of intelligent manufacturing machine

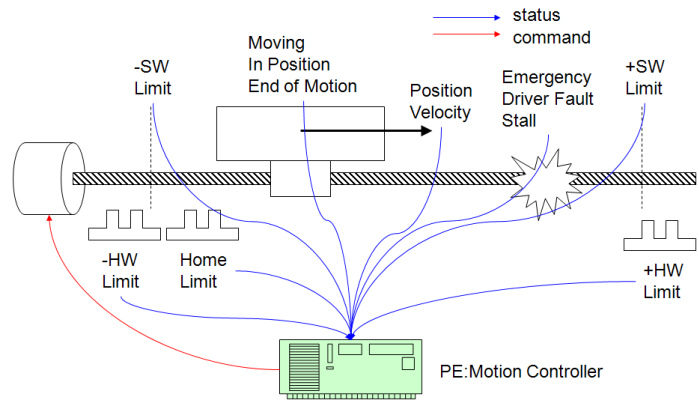

**Figure 2.** General function of motion controller

Member variables of VME, the SFs, store the current states of a MC. The VMEs are allocated in the computer memory and interface with individual physical MCs. Subroutines in an application can refer to the VME pointers when it needs checking motion status or moving an axis. The VME can queue or reject the commands from individual subroutines. An example of class header file for VME is shown as follows.

**Table 1.** An example of VME class coded by C++

| class CMotion<br>{<br>public:<br>    double m_dPos;<br>    double m_dVel;<br>    double m_dErr; | public:<br>    BOOL SJog(UINT nAx,double dV,double dA,double dJ);<br>    BOOL TJog(UINT nAx,double dV,double dA);<br>    BOOL SHome(UINT nAx,UINT nMode,double dV,double dV2,<br>        double dV3,double dA1,double dA2,double dJ1, double dJ2);<br>    BOOL THome(UINT nAx,UINT nMode,double dV,double dVb, |
|---|---|

```
    BOOL m_bHomLim;                    double dVf,double dA1,double dA2);
    BOOL m_bHwLimN;             BOOL SMove(UINT nAx,double dP,double dV,double dA1,
    BOOL m_bSwLimN;                    double dA2,double dJ1,double dJ2);
    BOOL m_bHwLimP;             BOOL TMove(UINT nAx,double dP,double dV,double dA1,
    BOOL m_bSwLimP;                    double dD2);
    BOOL m_bMoving;             BOOL UpdateStatus();
    BOOL m_bDecel;              void CloseBoard();
    BOOL m_bStall;              BOOL OpenBoard();
    BOOL m_bDrvFault;           void Initialize();
    BOOL m_bInPosition;       public:
    BOOL m_bCmdError;           CMotion();
    BOOL m_bEmergency;          virtual ~CMotion();
                              };
```

## 2.2 Machine vision

Machine vision performs image processing through frame grabbers(FGs) and can replace human eyes. Common tasks of FGs are capturing images, filtering, pattern recognition, finding edges, pixel operations and etc. The amount of mathematical calculation is larger than any other PEs, so computer resources are highly occupied during processing. Virtual vision equipment(VVE) can be generalized and subclassed by flags and commands. The command set of VVE is generalized into live video, capture image, import/export image, filter image, drawing figures on the image, pattern training/finding, edge finding, and image analysis and so on. The VVE can have SFs such as live images, captured images, active channels, acquisition state, inspection score, position of edge/pattern, processing time, and processed images. The VVE class is allocated in the computer memory, and subroutines refer the pointer like VME.

## 2.3 Data Acquisition

Commercial DAQ boards have IO ports for digital and analog. Digital signal has two states, the activated and the deactivated. If one of states has zero voltage, the other will have fixed voltage level such as +5V, +12V, and +24V. Digital input(DI) is such a signal from a peripheral equipment to PC and digital output(DO) is from PC to a peripheral equipment. Analog signal is converted into discrete voltage level in real time. The magnitude of those signals are usually measured by voltage base and sampled by time base. The resolution of voltage level is determined by the number of bits on signal buffer. AI is a signal transferred from analog port and AO is a signal forward out of analog port. So, the command of virtual DAQ equipment(VDE) handles the activation of DO channels, the output voltage of AO, and triggering. The flags of VDE represents current activation states in DIO and the voltage in AIO. VDE is also constructed and referred like above VEs.

## 2.4 Networking

Manufacturing machines may communicate with each other or be administrated by main server. The server, which is called factory automation system, schedules manufacturing progress and governs the work flow. Each machine should report current status such as product number, amount of work pieces, working time, system events, and tool usage. The server may give orders to start/stop process, change a tool, and event messages. The information is transferred and received in the from of CIM protocol. The virtual networking equipment(VNE) buffers packets, analyze network streams and converts the stream to SFs or execution code.

## 3. Software layers
### 3.1 Basic layer

Basic layer (BL) captures raw signals from PEs and updates flags by referring VEs. The signals of the PEs are sampled, converted and stored in VEs, which makes one cycle of monitoring. The layer is driven by window threads. For instance, a MC detects position, velocity, HW limit, and at al. The signals and status are captured by the command of VMC. The values are coupled with binary codes, so they are converted into real numbers by bitwise operations. The result of conversion is stored to flags in VMEs. A FG captures an image from a camera and executes image filtering by the commands of VME. The processing result and status of FGs are stored into VVE. AI signals are sampled and the voltage level at each port is updated into VDE. The SGL thread is executed until the program is terminated.

### 3.2 Logic layer

Logic Layer (LL) monitors the logical situation by combining the flags of the VEs. This layer converts the flags to logical flags and finds logical errors based on machine operating rules. This layer analyzes situations, detects errors and executes events. For example, when positive software limit(PSL) signal is detected, PSL flag in VME will be raised by the BL thread. LL notices the PSL flag, raises PSL error flag, stop the movement of mechanical parts, and notify the error to an active window. Signal interlock can be applied in this layer. When a user turns on a spindle, the LL raises spindle activation level(SAL). But one of the spindle feedback is off, logic layer refers to SAL and determines spindle error. Then, the spindle will be shut off for spindle protection. The voltage of VDE can be converted into the magnitude of vacuum pressure and checked for minimum level. When the pressure is out of threshold, LL cuts off the vacuum. The procedure is also executed by a window thread, and activated until the program is terminated. Fig. 3 explains concept of layer structure.

### 3.3 Display layer

The main function of display layer(DL) is to show the states of machine and information to a user. This layer displays the values measured by lower layers and updates window controls according to value of flags. When a window is ready to show, a DL thread is attached to the window. When the window is destroyed and new one is constructed, the threads for LL and BL continues working but the thread for DL will be destroyed. The working sequence can be explained as follows, for example. When a user touches a vacuum button in a window, the window will call a message function to activate vacuum by the commands of VEs. At the same time, the BL will update related flags and the LL check errors. The DL shows kinematical position or velocity, vacuum activation, and operating errors on the parent window. Fig. 4 shows the relation among VEs, windows and threads.

### 3.4 Process layer

Manufacturing sequence and algorithms are programmed into process layer(PL). PL is driven by a thread and performs manufacturing steps. The sequence is written by the commands of VEs. Working path and procedure are generated according to operating scenario. In wafer fabrication, the surface of a wafer can be inspected by a FG before and after processing. Then, the wafer is moved by tool path. This layer is created and attached to a manufacturing menu. The layer can be started, suspended, resumed and terminated by user's selection on the menu. The layer can notify its manufacturing events to the parent window when they are necessary.

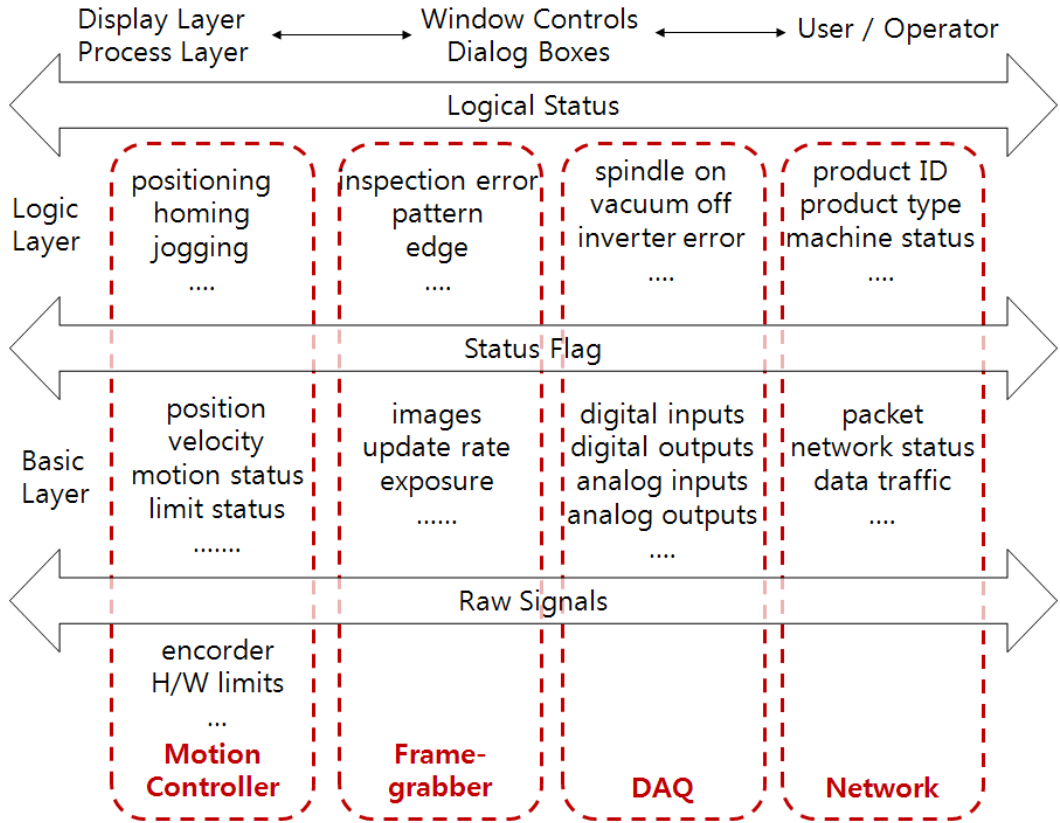

**Figure 3.** Thread and layers

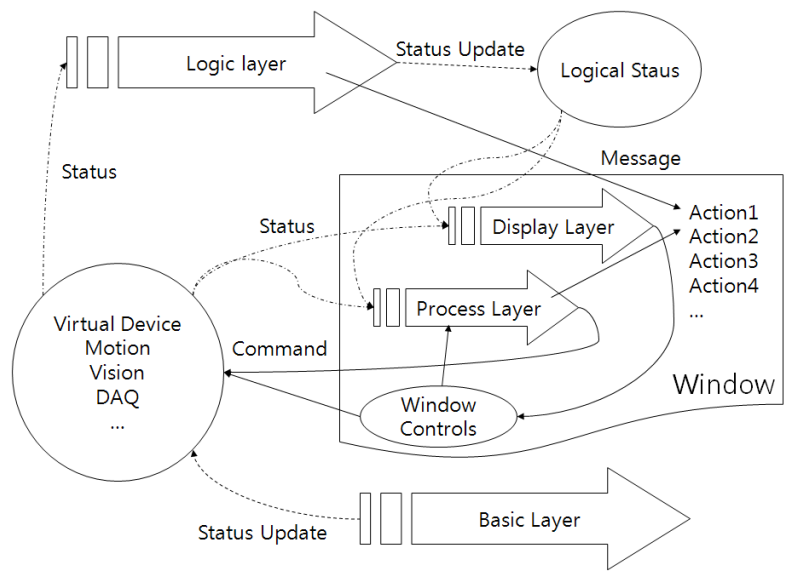

**Figure 4.** Data flow among windows, thread and objects

## 4. Simulation for efficiency
### 4.1 System load by PEs

Author names and affiliations are to be centered beneath the title and printed in Times New Roman 11-point, non-boldface type. Multiple authors may be shown in a two or three-column format, with their affiliations below their respective names. Affiliations are centered below each author name, italicized, not bold. Include e-mail addresses if possible. Follow the author information by two blank lines before main text.

The processing time and system load of industrial peripherals are much heavier than that of official or household ones. An ideal industrial computer can be assumed that the processing time of CPU, the system load by window display are infinitesimal, so most of the system load is occurred by PEs. Let N maximum capacity that a computer system can handle per unit time, $n_t$ and $n_w$ system load which are occupied by a thread and a window per unit time. System load by typical structure increases by the number of each window and thread, because the individual window and thread accesses to PEs. So, available capacity for $n_a$ is simply calculated as equation (1).

$$n_a = N - \sum_{i=1} n_{wi} - \sum_{j=1} n_{tj} \qquad (1)$$

Where, N, the maximum capacity of system resource, $n_t$, system load by threads per unit time, $n_w$, system load by windows per unit time, $n_a$, available capacity of system resource.

For an instance, each window accesses PEs to get machine status. Monitoring thread approach to data register on PEs before finding logical error. Process thread also tries sending commands and taking results in PEs. That is, system load can increase considerably when windows and threads are attached to program for a multi-tasking job. When $n_a$ reaches zero or negative, the computer will be down and manufacturing be stopped. Considering the proposed structure, the thread of BL fully acquire status of PEs and that of PL access PEs for launching command, so system load is governed by the two threads, which makes the possibility of system malfunction. Windows only refer the flags of VEs, so $n_w$ can be ignorable. Therefore, available capacity by the proposed architecture can be written as follows.

$$n_a = N - \sum_{j=1} n_{tj} \qquad (2)$$

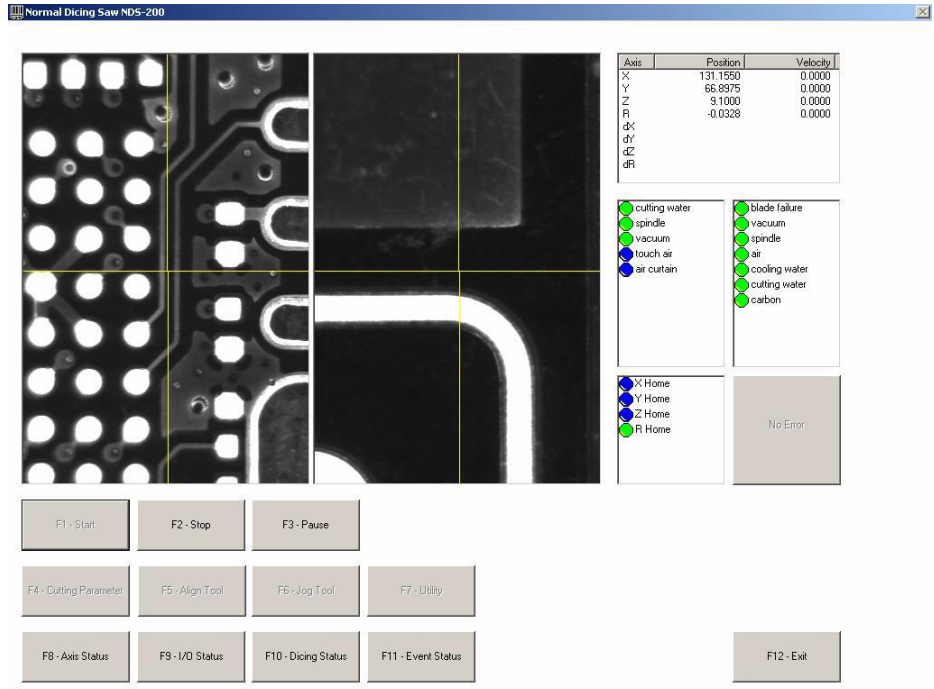

**Figure 5.** Window application for automatic wafer dicing

### 4.2 Simulation

It is difficult to define and measure system load exactly, but it is related with amount of computation and processing time. In this study, the number of access to PEs(NOA) is checked instead of the system load, because the load normally increases by NOA. In simulation, each command function performed nothing, but just counted NOA. The concept of software model is similar to Fig. 5. The software was constructed by C++ and libraries which were supplied by the board manufactures. Any open libraries were not applied to this software. The virtual software displays 2 channel of live vision, 7 mechanical signals in 4-axis, 8 DI's, and 5 DO's. Machine error is monitored for HW/SW limits, drive fault, emergency, spindle, blade, vacuum and coolant. Virtual manufacturing is consisted of 24 steps of automatic alignment and 144 steps of cutting movement in a dicing process. Detailed dicing process is explained in the next chapter. The typical model is event-based programming, which access to the hardware devices is conducted when necessary without considering the any architecture.

### 4.3 Result

Simulation result by the structure of is shown as Fig. 6 and the result by the proposed architecture is shown as Fig. 7. In each plot, x axis is dicing time, and y is NOA. NOA of MC decreased about 15% and that of DAQ especially fell down about 1/7. NOA of FG was much smaller than others, but it increased about 50 thousand. The dicing machine must monitor the status of motion during operation and immediately respond to exceptional status, so the NOA of MC decreased at small rate. It seems that the typical structure accesses VVE only when it needs, but the proposed structure always monitors its status. Table 2 shows the maximum and average of the NOAs in the Figs. 6 and 7. The NOA of the motion has the largest value but that of the vision does the smallest. The NOAs of the motion and DAQ were reduced but that of the vision increased. The variation rates of the maximum and the average in the motion, the vision and the total had almost same values. The NOA of the vision by the proposed architecture was almost same to that of the DAQ. The NOAs of the vision and the DAQ by the proposed had approximately 10% of that of the motion. The NOA of the vision by the proposed increased much because the access to the framegrabber was conducted in regular interval. The access to framegrabber in the typical method was done only when image processing begun. So, the status of the vision was unknown in the typical system. Although the NOA largely increased, the status of the vision can be monitored by the relatively small NOA in the total NOA, the increase can be tolerated. The total NOA by the proposed structure was reduced about 30 percent compared with typical one, and system load was lowered. This indicates that the load of processing in the operating software became lower and the possibility of the software shutdown decreased.

The model simulated in this paper is under assumption of small data generation which can be handled within millisecond level. In case of massive data generation, such as X-ray inspection machines, different model will be necessary including data storage and transfer.

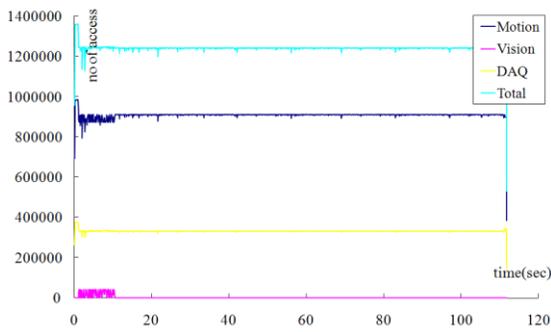
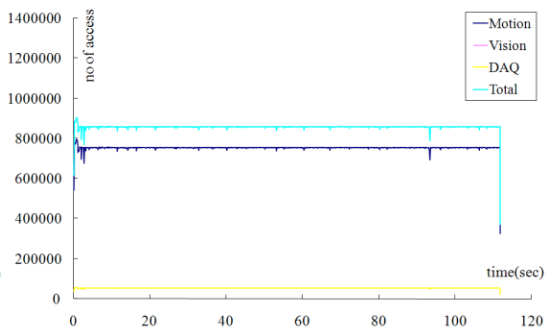

Figure 6. NOA to PEs in typical architecture    Figure 7. NOA to PEs in proposed architecture

Table 2. Comparing the NOAs by the typical and proposed architecture

| devices | Typical | | Proposed | | Difference | | Rate (%) | |
|---|---|---|---|---|---|---|---|---|
| | Max | Avg | Max | Avg | Max | Avg | Max | Avg |
| Motion | 984768 | 908376 | 792872 | 752320 | -191896 | -156056 | -19 | -17 |
| Vision | 42409 | 1849 | 54619 | 51893 | 12210 | 50044 | 29 | 2706 |
| DAQ | 375178 | 330952 | 54800 | 51883 | -320378 | -279068 | -85 | -84 |
| Total | 1359954 | 1241178 | 902291 | 856097 | -457663 | -385080 | -33 | -31 |

## 5. Application for wafer dicing
### 5.1 Dicing process

Dicing is the wafer cutting process, having a wafer separated to chips. A wafer is placed on a table and aligned along the direction of a cutting blade. The blade made of industrial diamond is rotated at 10,000rpm - 60,000rpm by an air spindle. The blade is lowered until it contacts with a wafer, and the wafer is pushed along blade direction. When the cutting have finished, the blade is positioned above next cutting position, then the cutting procedure is repeated. Fig. 8 shows the basic concept of dicing process and tool path. After finishing the cutting, the quality of cutting lines can be checked with machine vision.

### 5.2 Dicing process

Automatic dicing machine interfaces with a MC, a FG, a DAQ board and a network card. The machine has XYZƟ-four axes. XYZ axes are for translational or linear movement. Those axes can be composed of linear guide-ball screws or linear motors. Ɵ axis is for rotational movement, which is consisted of a harmonic drive and a motor. The MC always watches motion status on each axis and drive motors. Misalignment is inspected by FGs [13] and the kinematical position is compensated. The machine can change the activation level of a spindle, coolant, vacuum and lamps during processing. The machine monitors error and shows the status in a screen. So, the proposed architecture can be applied to develop a software for the dicing machine. The specification of the automatic dicing machine in this study is shown in Table 3.

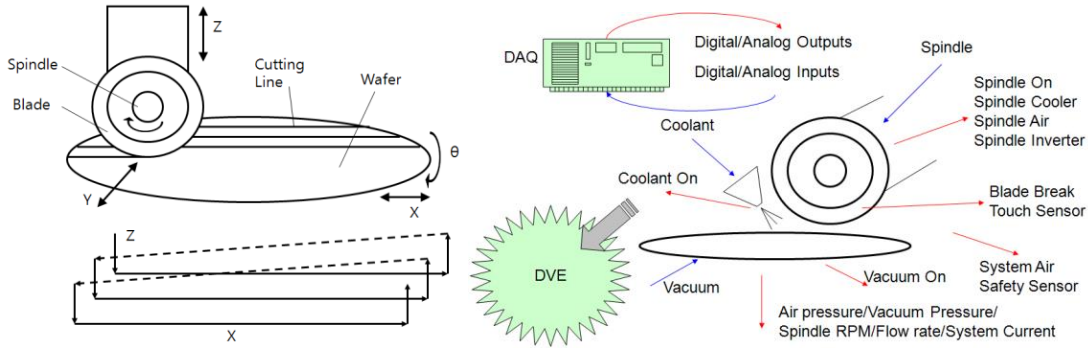

**Figure 8.** Working path for the dicing    **Figure 9.** Peripheral equipments and signals in dicing machine

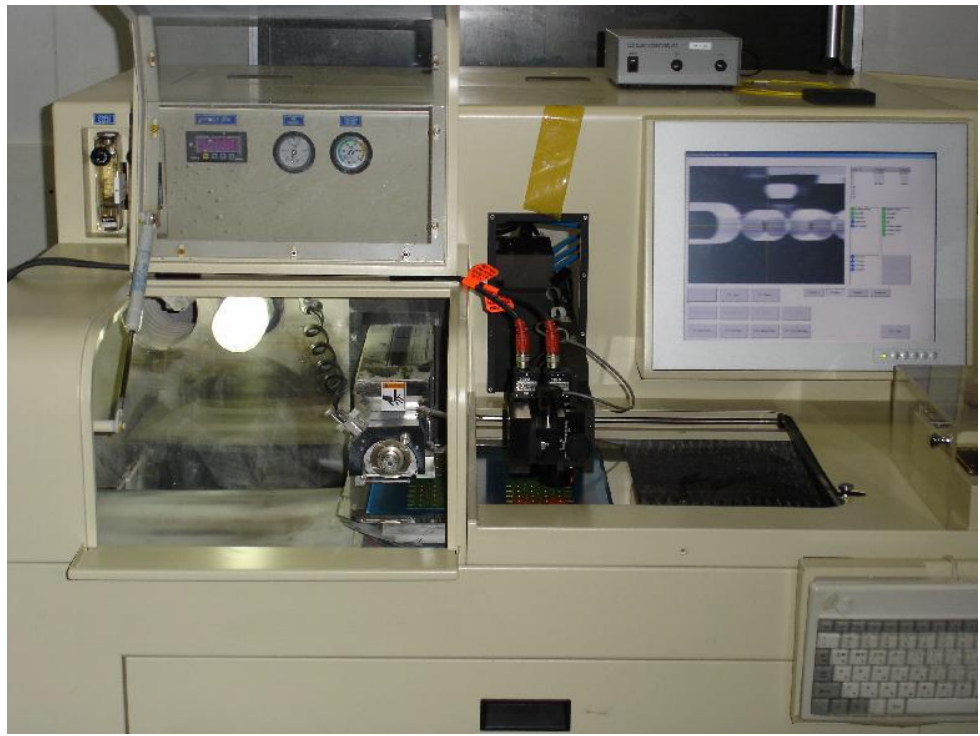

**Figure 10.** Photo of automatic wafer dicing

Table 3. Specification of the dicing machine

| devices | Assemblies / bender | amounts |
|---|---|---|
| Motion control | Driver / Mitsubishi | 4 |
| | Controller / Ajin-Extek | 2 |
| | Motor & LM Guide / TSK | 3 |
| | Harmonic Drive / Sam-Ick) | 1 |
| Vision | Famegrabber / Cognex | 1 |
| | Camera / Hitachi | 2 |
| | Light Source / BaekDu | 1 |
| DAQ | DAQ Boards / Ajin-Extek | 2 |
| | Digital Input / Ajin-Extek | 32 |
| | Digital Output / Ajin-Extek | 32 |

### 5.3 Software

Fig. 5 shows a main window for conducting automatic dicing process, built by VC++. Each PE was coded into VEs complied into DLL. The VEs were allocated in the memory and the pointers were referred from threads, message functions and tool windows. BL monitored the status of the MC, the FG, and the DAQ. LL watched errors on axes, spindle, air, vacuum and blade. Interlocks were set on spindle, vacuum and coolant. When an error was detected by LL, an event message was posted to the main window and activated error button. The error was cleared only when the error button was touched. The dicing process was programmed on the PL. The PL aligned wafers, controlled dicing tool path, and determined the activation of coolants/lamps. The thread could be stared, stopped, suspended and resumed by touching buttons in the window. An operator could input dicing parameters, aligns a wafer, makes a jog movement, and activate peripheral equipments on the tool windows. DL showed two images from cameras, kinematical position, velocity, spindle, vacuum, cutting water, blade sensors and home limits. Each tool window had own DL to show various information. Fig. 10 is the picture of dicing machine which the software is developed for. The dicing machine was installed at Samsung Electro Mechanics for manufacturing Blue Tooth chips and handles 60-80 thousand sheets of PCB a month.

### 6. Conclusion

The software structure for operating semiconductor manufacturing machine with industrial equipments was proposed. The physical equipments were abstracted and coded into virtual equipments using C++ class. The virtual equipments were allocated in the computer memory and he subroutines referred to the pointers of virtual equipments, but direct access to physical equipments was not allowed. The machine was managed by four threads, such as basic layer, logic layer, display layer and process layer. Each layer functioned signal detection, finding logical errors, displaying information and executing manufacturing sequence. The efficiency of the typical and the proposed structure was simulated. In the result, the number of access to PEs showed that system load, the resource of a control PC, could be reduced about 30%. The architecture was applied to operating software for an automatic dicing machine which is currently running in a factory.

### Acknowledgement

This work was funded and supported by AM Technology (http://www.amtechnology.co.kr) and the authors are grateful to them for supplying us with a dicing machine. The next work is massive data manipulation in an image-based inspection machine. The machine also interfaces with multiple signal boards but the size of data reaches tens of GB and it frequently makes confliction to compiling codes by mixing libraries of the boards. So we will consider massive data acquisition, high speed transfer, accelerated processing and avoiding the library confliction in the next study.

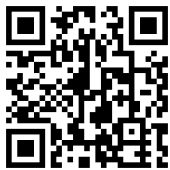